# Mechanical Properties of Pentagraphene-based Nanotubes: A Molecular Dynamics Study

J. M. de Sousa[1,2], A. L. Aguiar[2], E. C. Girão[2], Alexandre F. Fonseca[1], A. G. Sousa Filho[3], and Douglas S. Galvao[1]

[1]Applied Physics Department and Center of Computational Engineering and Science, University of Campinas - UNICAMP, Campinas-SP 13083-959, Brazil.

[2]Departamento de Física, Universidade Federal do Piauí, Teresina-PI, 64049-550, Brazil.

[3]Departamento de Física, Universidade Federal do Ceará, Fortaleza-CE, 60445-900, Brazil.

ABSTRACT

*The study of the mechanical properties of nanostructured systems has gained importance in theoretical and experimental research in recent years. Carbon nanotubes (CNTs) are one of the strongest nanomaterials found in nature, with Young's Modulus (YM) in the order 1.25 TPa. One interesting question is about the possibility of generating new nanostructures with 1D symmetry and with similar and/or superior CNT properties. In this work, we present a study on the dynamical, structural, mechanical properties, fracture patterns and YM values for one class of these structures, the so-called pentagraphene nanotubes (PGNTs). These tubes are formed rolling up pentagraphene membranes (which are quasi-bidimensional structures formed by densely compacted pentagons of carbon atoms in $sp^3$ and $sp^2$ hybridized states) in the same form that CNTs are formed from rolling up graphene membranes. We carried out fully atomistic molecular dynamics simulations using the ReaxFF force field. We have considered zigzag-like and armchair-like PGNTs of different diameters. Our results show that PGNTs present YM ~ 800 GPa with distinct elastic behavior in relation to CNTs, mainly associated with mechanical failure, chirality dependent fracture patterns and extensive structural reconstructions.*

INTRODUCTION

Recently, a new carbon allotrope named pentagraphene (PG) was theoretically proposed [1]. PG is similar to graphene, but instead of having hexagonally arranged carbon atoms, it has pentagonal ones. PG exhibits interesting properties, such as intrinsic geometric wrinkles and negative Poisson's ratio [1,2]. As carbon nanotubes can be considered as graphene sheets rolled up into cylindrical topology, a natural question is whether PG-based nanotubes (PGNTs) could exist. A recent work [3] using molecular dynamics methods showed that PGNTs are stable and under axial deformations present plastic characteristics and structural transitions. However, a more detailed and comprehensive PGNT study is still needed. In this work, we carried out extensive fully atomistic reactive molecular dynamics (MD) simulations to investigate the structural and fracture patterns of different PGNTs (diameters and chiralities, see Figure 1). The MD simulations were carried out using the reactive interatomic potential ReaxFF [4], as implemented in LAMMPS code [5].

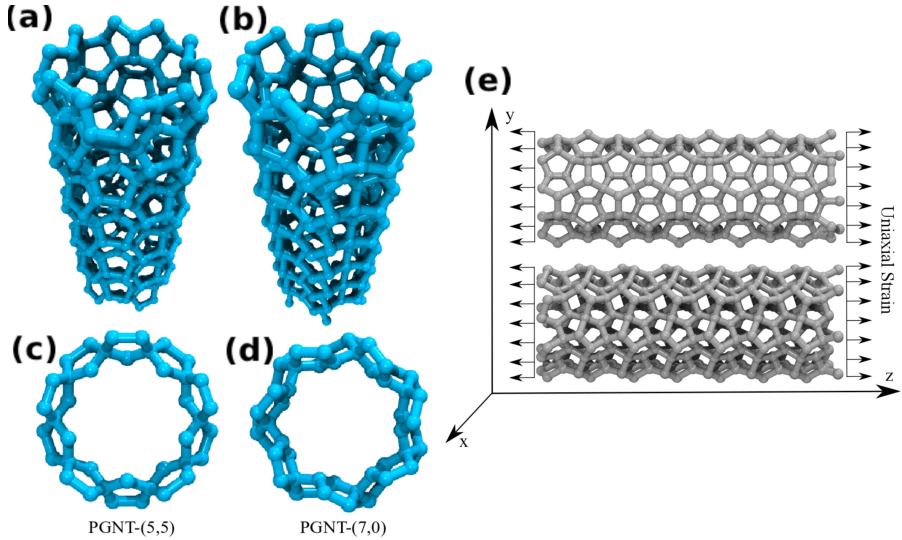

Figure 1: Atomic model for pentagraphene-based carbon nanotubes (PGNTs). (a) PGNT-(5,5) with a armchair-like edge and diameter of 8.20 Å; (b) PGNT-(7,0) with a zigzag-like edge and diameter of 9.10 Å; (c) and (d) a perpendicular view of the PGNTs with edge armchair-like and zigzag-like respectively; (e) Atomic model indicating the direction (z axis) of the applied uniaxial strain for PGNT-(5,5) (top panel) and PGNT-(7,0) (bottom panel).

**METHODS**

We investigated the dynamical, structural, fracture patterns and Young's modulus of armchair-like and zigzag-like (edges) PGNTs of different diameters. These properties were obtained from fully atomistic molecular dynamics (MD) simulations using the reactive potential ReaxFF [4], as available in the LAMMPS [5] code. ReaxFF is parameterized directly from calculations based on first principles and compared with experimental values. For hydrocarbon systems, the deviations between the predicted values and the experimental data for the heat of formation values of unconjugated and conjugated systems are approximately 2.8 and 2.9 Kcal/mol, respectively [6]. ReaxFF allows the description of break and formation of covalent chemical bonds as a function of the bond order values, thus being ideal for the study of the mechanism of fracture in nanostructured systems. The structures considered in this study were zigzag-like (**n,0**) and armchair-like (**n,n**) PGNTs nanotubes with **n** varying from 5 up to 10. This corresponds to tube diameters from $5.8\ \text{Å} - 11.6\ \text{Å}$ (zigzag-like) and $8.2\ \text{Å} - 16.2\ \text{Å}$ (armchair-like). We considered tube lengths of approximately $26\ \text{Å}$. These tubes are representative of PGNT families.

$$Y = \frac{\sigma_{ii}}{\varepsilon_i}$$

$\sigma_{ii}$ $\varepsilon_i$ $i$

In order to assure zero initial stress before the start of the uniaxial stretching dynamics, we turn null stress at the tube edges with the use of the ensemble *isothermal-isobaric* (NPT) [8]. Also before starting to stretch the tubes along the z-direction (Figure 1), they were thermalized at 300 K using an ensemble canonical (NVT) and with a Nosé-Hoover thermostat [7], as implemented in the LAMMPS code.

$\sum_k^N m_k v_k v_k$ $\sum_k^N m_k r_k f_k$
$\sigma_{ji} = \frac{}{A} \quad \frac{}{A}$

The structural changes resulting from the stretching dynamics are updated at each molecular dynamics time steps runs of *0.05 fs*, which is sufficient time for the entire nanotube to be equilibrated before the next simulation step. The applied tensile strain rate is simulated by gradual increasing of the simulation box at a rate of $10^{-6}$/fs. The structural deformations are monitored through the Young's Modulus and von Mises stress (calculated from the virial stress tensor) values [9]. The stretching dynamics is continuously applied until the tubes reach permanent deformations and/or structural failure (fracture).

$1.53\,\text{Å}$ $1.57\,\text{Å}$
$1.535\,\text{Å}$ $1.555\,\text{Å}$

## DISCUSSION

The carbon-carbon bond-lengths from the ReaxFF optimized geometries were 1.53 Å and 1.57 Å, which are in very good agreement with DFT results published in the literature [1], 1.54 Å and 1.56 Å, respectively.

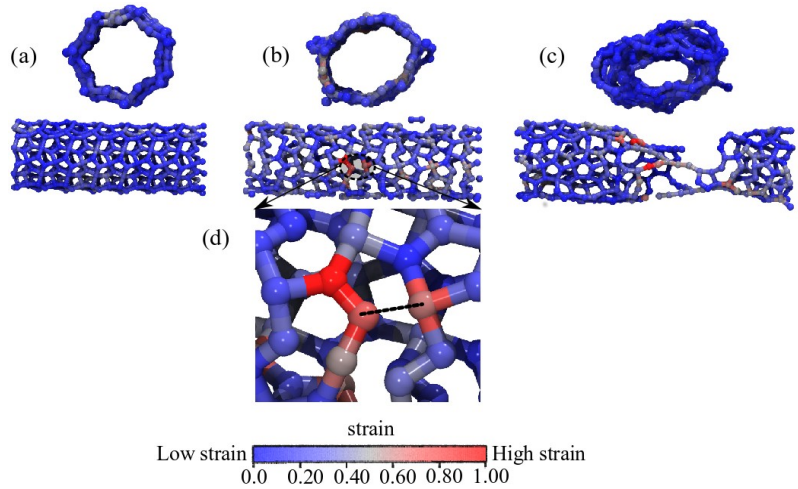

Figure 2: MD snapshots of PGNT zigzag-like (7,0) at different stretching stages. (a) Initially equilibrated at 300K and at 0% strain. (b) Tensioned at 13% strain, where we can already observe some broken bonds (c) Almost completely fractured at 34% strain and; (d) Close up of figure 2(b) showing one of the broken bonds (black dotted line). The horizontal colored bar indicates von Mises stress values.

We next analyzed the results of the applied uniaxial strain. In Figure 2, we present representative MD snapshots of the stretching dynamics at different stages (from the unstressed states up to almost complete fracture) for the case of PGNT(7,0). As we can see from Figures 2b and 2d, at 13% strain, we can already observe some broken bonds. As expected, the bonds that are approximately parallel to the direction of the applied uniaxial strain (z direction) are the first to break, as shown in the enlarged box in (Figure 2d), where the black dotted line indicates one of these broken bonds. At 34% strain, the structure is almost completely fractured. The critical strain for this case was $\sigma_c$= 18%. The horizontally colored bar in Figure 2 indicates the von Mises stress concentrations, where blue and red colors indicate low and high strain regimes, respectively.

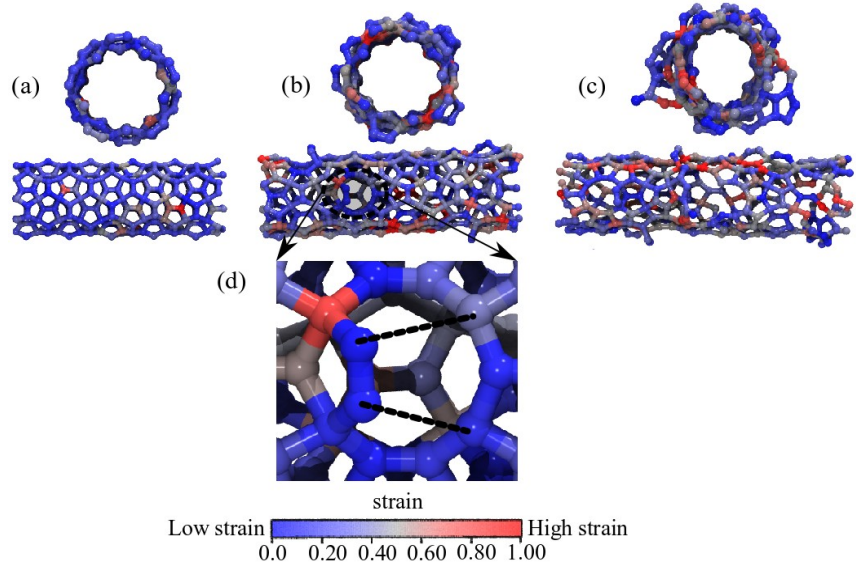

Figure 3: MD snapshots of PGNT armchair-like (7,0) at different stretching stages. (a) Initially equilibrated at 300K and at 0% strain. (b) Tensioned at 9% strain, where we can already observe some broken bonds (c) Almost ompletely fractured at 23% strain and; (d) Enlarged fractured region, the black dotted lines indicate some of the broken bonds. The horizontal colored bar indicates von Mises stress values.

In Figure 3, we present the corresponding results for the PGNT(7,0). As we can see from Figures 2 and 3, the main features are similar for the zigzag-like and armchair-like tubes. Themain difference is that armchair-like tubes are fractured at a much earlier stage (23% in comparison to 34%) than the corresponding zigzag-like ones. This is a direct consequence of the structural bond alignments in relation to the direction of the applied tension (Figure 3d), which results in a more brittle behavior.

The PGNT Young's modulus values were estimated from the linear regime of the stress-strain curves (see Figure 4). The fracture patterns were obtained from MD trajectories and using the von Mises stress tensor [9]. Our results show that PGNT armchair-like and zigzag-like have, on average, an elasticity modulus of 800 GPa, which is 36% lower that the corresponding values for CNTs (~1250GPa [10]).

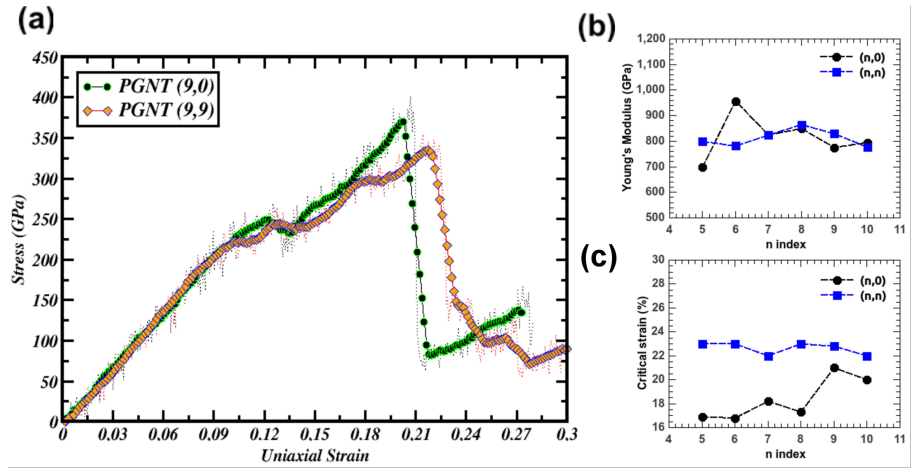

Figure 4: (a) Stress strain curve for PGNT with chirality zigzag-like (9,0) and armchair-like (9,9). For the PGNT zigzag-like, the critical strain is 20.5% and the PGNT armchair-like is 22.6%. (b) Young's Modulus values for zigzag-like and armchair-like PGNTs. (c) Critical strain $\sigma_c$ values for zigzag-like and armchair-like PGNTs.

## CONCLUSION

We have investigated the mechanical properties and fracture patterns of armchair and zigzag-like pentagraphene nanotubes (PGNTs). Pentagraphene is a quasi-2D (buckled) carbon allotrope composed of pentagons (and not hexagons, like in graphene). Our results show that PGNT fracture patterns is chirality dependent. This is due to the different number of carbon-carbon bonds aligned with the direction of the applied tension (see Figures 2 and 3), which results in armchair tubes breaking at earlier stages than the corresponding zigzag-like ones. The estimated PGNT Young's modulus values for the structures considered here were, on average, 800 GPa, which is 36% lower than conventional carbon nanotubes.

## ACKNOWLEDGEMENTS

This work was supported in part by the Brazilian Agencies CAPES, CNPq and FAPESP. The authors thank the Center for Computational Engineering and Sciences at Unicamp for financial support through the FAPESP/CEPID Grant #2013/08293-7. AFF acknowledges support from FAPESP grant #2016/00023-9. AGSF, ECG and JMS acknowledge support from CAPES through the Science Without Borders program (A085/2013)